\documentclass[a4paper,iopart]{jpconf}
\usepackage{graphicx}
\usepackage{amsmath}
\begin{document}

\title{Implications of finite one-loop corrections for see-saw neutrino
  masses}

\author{D Aristizabal Sierra}

\address{IFPA, Dep. AGO, Universite de Liege, Bat B5, Sart Tilman
  B-4000 Liege 1, Belgium}

\ead{daristizabal@ulg.ac.be}

\begin{abstract}
  In the standard seesaw model, finite corrections to the neutrino
  mass matrix arise from one-loop self-energy diagrams mediated by a
  heavy neutrino. We discuss the impact that these corrections may
  have on the different low-energy neutrino observables paying special
  attention to their dependence with the seesaw model parameters. It
  is shown that sizable deviations from the tri-bimaximal mixing
  pattern can be obtained when these corrections are taken into
  account.
\end{abstract}

\section{Introduction}
\label{sec:intro}
One of the most simplest realizations of the dimension five effective
operator ${\cal O}_5\sim LLHH$ is certainly given by the {\it
  standard} seesaw model
\cite{Minkowski:1977sc,Yanagida:1979as,Mohapatra:1979ia,
  GellMann:1980vs,Schechter:1980gr}. By extending the standard model
Lagrangian with fermionic electroweak singlets (right-handed (RH)
neutrinos for brevity) light neutrino masses are generated by the
exchange of the new states via a tree-level realization of ${\cal
  O}_5\sim LLHH$. An appealing feature of this approach is that the
smallness of neutrinos masses is due to the suppression induced by
the new heavy states.

The seesaw Lagrangian spans an 18-dimensional parameter space
determined by 6 CP phases and 12 real parameters. This parameter space
is partially constrained by low-energy neutrino data
\cite{Schwetz:2008er}: these measurements -in principle- yield 9
constraints given by 3 light neutrino masses, 1 Dirac and 2 Majorana
CP phases and 3 mixing angles. The remaining 9-dimensional parameter
space is not constrained at all implying a region in parameter space
leading to the correct low-energy observables can not be uniquely
established.

In order to find those ``spots'' that are consistent with data
typically scans of the seesaw parameter space are done. These scans
rely on parametrizations of the seesaw parameters (Yukawa couplings)
in such a way that instead of using low-energy data as an output
criteria it is used as an input \footnote{For a discussion of
  parametrizations of the seesaw see
  \cite{Davidson:2004wi}.}. In this procedure light neutrino
masses are taken to be entirely determined by the tree-level mass
matrix with the Yukawa couplings restricted to be perturbative
\cite{Casas:2010wm}. This an accurate approximation as far as the one-loop
contributions are not too large. However, depending on the parameter
choice, regions in parameter space exist in which the one-loop
contributions, and in particular the finite ones, are sizable
\cite{AristizabalSierra:2011mn,AristizabalSierra:2011gg}. Thus
implying that the perturbativity of the Yukawa couplings though being a
proper condition it is not sufficient to guarantee a reliable scan of
the seesaw parameter space.

Here we will discuss the importance of the finite one-loop corrections
for seesaw neutrino masses. We will present the general formulas for
the self-energy diagrams that determine them, showing how can they
affect the different entries of the light neutrino mass matrix. As an
implication of these corrections we will analyze the impact they may
have on the leptonic mixing angles by assuming that the tri-bimaximal
mixing (TBM) is realized at the tree-level. We will closely follow
reference \cite{AristizabalSierra:2011mn}.
\begin{figure}
  \centering
  \includegraphics[width=7cm,height=2cm]{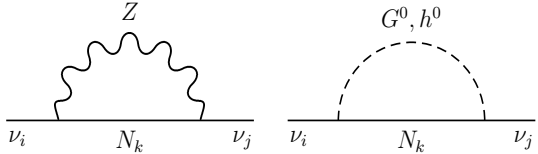}
  \caption{Self-energy diagrams accounting for $\pmb{m_\nu}^{\text{(1-loop)}}$.}
  \label{fig:slef-energy-diagrams}
\end{figure}
\section{One-loop finite corrections}
The new states being $SU(2)\times U(1)$ singlets do not couple to
standard model gauge bosons. Thus, in the basis in which the Majorana
RH neutrino mass matrix is diagonal the interactions induced by the
presence of these states, that for concreteness we assume to be three,
are described by the following Lagrangian:
\begin{equation}
  \label{eq:seesaw-Lag}
  -{\cal L}=-i\bar N_{R_i}\,\gamma_\mu\partial^\mu N_{R_i}
  + \tilde\phi^\dagger\bar N_{R_i}\lambda_{ij}\ell_{Lj}
  + \frac{1}{2}\bar N_{R_i} C M_{R_i} \bar N_{R}^T
  + \mbox{h.c.}
\end{equation}
Here $\ell_L$ are the lepton electroweak doublets, $\phi^T=(\phi^+
\phi^0)$ is the Higgs $SU(2)$ doublet, $M_{R_i}\equiv M_i$ are the RH
neutrino massses, $C$ is the charge conjugation operator and
$\pmb{\lambda}$ is a $3\times 3$ Yukawa matrix in flavor space.
Assuming $M_R\gg v$ (with $v\simeq 174$ GeV) the effective neutrino
mass matrix is obtained once the full $6\times 6$ neutral fermion mass
matrix is diagonalized:
\begin{equation}
  \label{eq:seesaw-formula-tree-level}
\pmb{m_\nu}^{\mbox{\tiny(tree)}}=
-v^2\pmb{\lambda}^T\,\pmb{\hat M_R}^{-1}\,\pmb{\lambda}\,.
\end{equation}
\begin{figure}
  \centering
  \includegraphics[width=14cm,height=9cm]{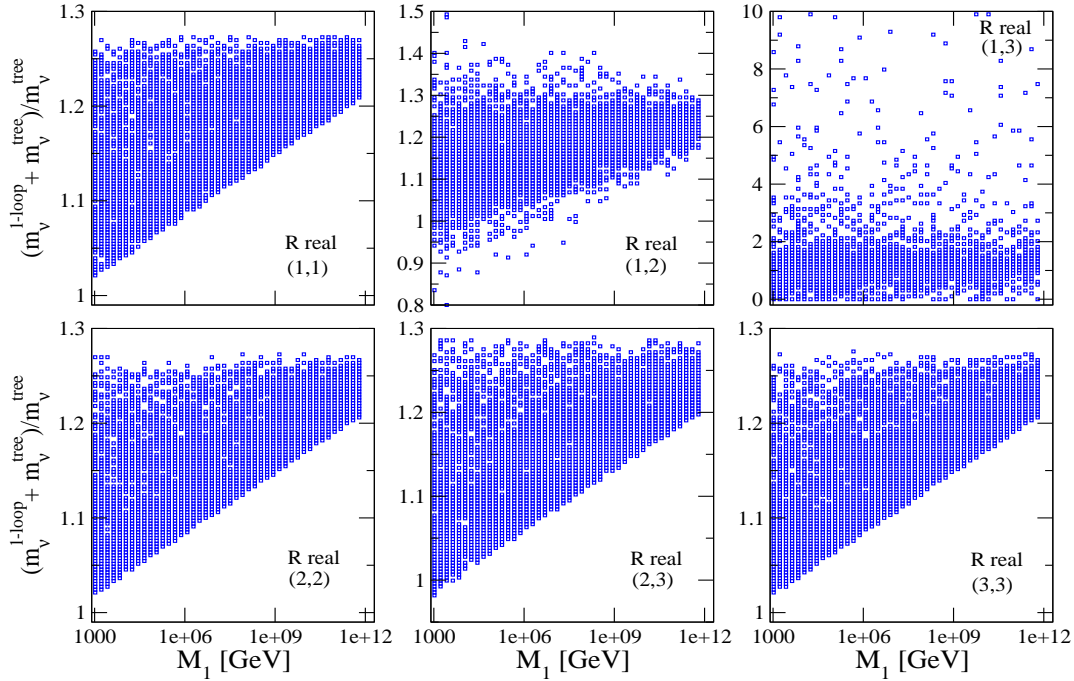}
  \caption{Ratio between the 1-loop and tree level elements of the
    light neutrino mass matrix as a function of the lightest RH
    neutrino mass $M_1$. A normal light neutrino mass spectrum has
    been assumed.}
  \label{fig:harrayRR}
\end{figure}
As pointed out in references \cite{Pilaftsis:1991ug,Grimus:2002nk}
finite corrections to the above matrix arise from the one-loop
self-energy Feynman diagrams depicted in figure
\ref{fig:slef-energy-diagrams}. These corrections remove the $3\times
3$ null matrix in the left-upper block of the full $6\times 6$ fermion
mass matrix. Thus, their finiteness can be understood as due to the
fact that there are no counterterms that could be used to absorb
possible divergences arising from them.  Full details of the
calculation can be found in the appendix of reference
\cite{AristizabalSierra:2011mn}, the result reads
\begin{equation}
   \label{eq:deltaML}
   \pmb{m_\nu}^{\mbox{\tiny(1-loop)}}=
   v^2\pmb{\lambda}^T \pmb{\hat M_R}^{-1}
   \left\{\frac{g^2}{64 \pi^2 M_W^2}
     \left[
       m_h^2\ln\left(\frac{\pmb{\hat M_R}^2}{m^2_h}\right)
       +
       3 M_Z^2\ln\left(\frac{\pmb{\hat M_R}^2}{M^2_Z}\right)
     \right]\right\}\pmb{\lambda}\,.
\end{equation}
Some words are in order regarding this result. The correction is not
suppressed with respect to the tree-level result by additional factors
of $v\,\pmb{\lambda}\,\pmb{M_R}^{-1}$. Thus, it is expected to be
smaller than the tree-level mass term solely by a factor of order
$(16\pi^2)^{-1}\ln(M_R/M_Z)$, implying it might have sizable effects.
The logarithms involving the RH neutrino masses imply the correction
is not proportional to the tree-level mass matrix in
(\ref{eq:seesaw-formula-tree-level}), unless the RH neutrino mass
spectrum is degenerate. Accordingly, barring this case the correction
can modified the tree-level result in a sizable way.

By using the Casas-Ibarra parametrization \cite{Casas:2001sr}, namely
\begin{equation}
  \label{eq:c-i-p}
  \pmb{\lambda}=\frac{\sqrt{\pmb{\hat M_R}}\,\pmb{R}\,\sqrt{\pmb{\hat
        m_\nu}} \,\pmb{U}^\dagger}{v}\,,
\end{equation}
where $\pmb{R}$ is a general complex orthogonal matrix, we scan the
parameter space assuming a normal hierarchical light neutrino spectrum
and a real $\pmb{R}$. The results for the elements of the mass matrix
are displayed in figure \ref{fig:harrayRR}. It can be seen that
even in this conservative case \footnote{Note that under this assumption
  three out of the six CP physical phases are taken to be zero.}
the corrections can be of ${\cal O}\sim 30\%-50\%$, the exception being
the 13 entry for which the contribution is far more large due to the
vanishing of this entry at tree-level at around $\theta_{13}\sim
5^\circ$.

\begin{figure}
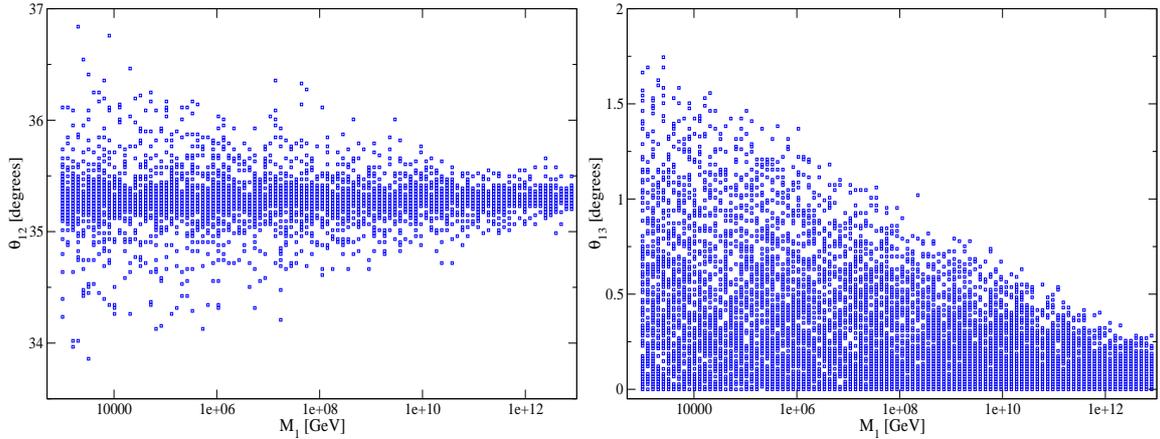

  \centering
  \includegraphics[width=7.5cm,height=5.8cm]{tribit12.eps}
  \includegraphics[width=7.5cm,height=5.8cm]{tribit13.eps}
  \caption{Deviations from the TBM pattern induced by the the one-loop
    finite corrections as a function of the lightest RH neutrino
    mass. The left panel (right panel) shows the results for the solar
    (reactor) neutrino mixing angle.}
  \label{fig:mixing-angles}
\end{figure}

Although the corrections in (\ref{eq:deltaML}) can sizably modify the
neutrino mass matrix not necessarily may have such an effect in
low-energy observables. For example, if the modification of the
neutrino mass matrix elements are correlated the effects on the
neutrino mixing angles could be negligible as they are calculated from
ratios of the mass matrix entries. In order to see whether this is the 
case we have assumed an input {\it ansatz} for the neutrino mixing angles
that we have chosen to be the TBM pattern:
\begin{equation}
  \label{eq:tbm}
  \theta_{23}=45^\circ\,,\quad \theta_{12}=33.3^\circ\,,\quad 
  \theta_{13}=0^\circ\,.
\end{equation}
Assuming $\pmb{R}$ to be real, with these values, we calculate the
corresponding Yukawa couplings from (\ref{eq:c-i-p}) for RH neutrino
masses ranging from $[10^{3},10^{12}]\,$ GeV.  With these Yukawa
couplings we proceed to calculate the full effective mass matrix
$\pmb{m_\nu}^\text{eff}=\pmb{m_\nu}^\text{(tree)}+\pmb{m_\nu}^\text{(1-loop)}$
and the corresponding eigenvectors from which we extract the mixing
angles. The results are displayed in figure \ref{fig:mixing-angles}
for the solar and reactor angle as well. It can be seen that sizable
deviations from the TBM pattern are more pronounced for small values,
${\cal O}(\mbox{TeV})$, of $M_1$: in models with a light lightest RH
neutrino there is more room for large hierarchies in the RH neutrino
spectrum and accordingly the mismatch between the tree-level effective
matrix and the finite one-loop correction (\ref{eq:deltaML}), due the
logarithmic functions, is larger. As in the previous case this example
can be regarded as conservative and larger deviations could be
expected in the general case of complex $\pmb{R}$ (see
ref. \cite{AristizabalSierra:2011mn}).
\section{Conclusions}
\label{sec:conc}
In the seesaw model finite one-loop corrections arise from self-energy
diagrams mediated by a heavy RH neutrino. We have analyzed the
importance these corrections may have for the different entries of the
effective light neutrino mass matrix finding the deviations from the
tree-level result to be quite relevant. We also studied the impact of
these corrections on low-energy observables by using the TBM mixing
pattern as an input {\it ansatz}. As our results show sizable
deviations from this pattern can be expected. In summary our findings
prove one-loop finite corrections to seesaw neutrino masses are
important and thus should be included in any analysis of the seesaw
parameter space.  \bibliography{neutrino}
\end{document}